\documentclass[12pt]{article}
\usepackage[left=1in,right=1in,top=1in,bottom=1in]{geometry}

\usepackage[normalem]{ulem}
\usepackage[table]{xcolor}
\usepackage{longtable}
\usepackage{xltabular}
\usepackage{tabulary}
\usepackage{amssymb}
\usepackage{amsthm}
\usepackage{amsmath}
\usepackage{cancel}
\usepackage{gensymb}

\usepackage[colorlinks,allcolors=blue,bookmarks=false,hypertexnames=true]{hyperref}
\usepackage{tikz}
\usetikzlibrary{automata,positioning,arrows,shapes.geometric}
\usepackage{multirow}
\usepackage{commath}
\usepackage{verbatim}
\usepackage{tikz-3dplot}
\tdplotsetmaincoords{60}{150} 
\tdplotsetrotatedcoords{0}{0}{0} 
\graphicspath{{./figures/}}
\usepackage{xcolor}
\definecolor{dark-violet}{RGB}{148,0,211}
\usepackage{subcaption}
\usepackage{tikz}
\usetikzlibrary{positioning}
\usepackage{blindtext}
\usepackage{multirow}
\usepackage{enumitem}
\usepackage{chemformula}
\usepackage{pstricks}
\usepackage{import}
\usepackage{caption}

\usepackage{filecontents}

\usepackage[sorting = none, backend = bibtex, style=numeric-comp, url = false]{biblatex}
\usepackage{authblk}

\usepackage[section]{placeins}

\title{\textbf{Designing a thermodynamically stable and intrinsically ductile refractory alloy}}
\author[1]{Sufyan M. Shaikh}
\author[1, 2]{B. S. Murty}
\author[*1,3]{Satyesh K. Yadav}
\affil[1]{Department of Metallurgical and Materials Engineering, Indian Institute of Technology Madras, Chennai, 600036, Tamil Nadu, India}
\affil[2]{Materials Science \&  Metallurgical Engineering, Indian Institute of Technology Hyderabad, Kandi, 502285, Telangana, India}
\affil[3]{Center for Atomistic Modeling and Materials Design, Indian Institute of Technology Madras, Chennai, 600036, Tamil Nadu, India}
\date{*Corresponding author: \href{mailto:satyesh@iitm.ac.in}{satyesh@iitm.ac.in}\\Contributing authors: \href{mailto:sufyanshk@gmail.com}{sufyanshk@gmail.com}; \href{mailto:bsm@iith.ac.in}{bsm@iith.ac.in}}

\begin{document}
\maketitle
\begin{abstract}
Developing ductile refractory BCC alloys has remained a challenge. The intrinsic ductility (D) of an alloy is the ratio of surface energy ($\gamma_s$) and unstable stacking fault energy ($\gamma_{usfe}$). Lowering the valence electron concentration has been shown to improve the intrinsic ductility of refractory alloys. However, Re has been widely used to ductilize W, contrary to the low valency criteria suggested in the literature. Here we use density functional theory to calculate the enthalpy of formation, $\gamma_{usfe}$ and $\gamma_s$  of Group IV, V, VI elements and their 25 equiatomic binary alloys in BCC crystal structure. We found that positive enthalpy leads to a considerable reduction in $\gamma_{usfe}$ compared to composition averaged value, resulting in improved intrinsic ductility. Enthalpy is maximum at the equiatomic concentrations indicating the highly repulsive interaction between the alloy constituents and vicer-versa. We found that the repulsive interaction between the alloy constituents leads to a reduction in $\gamma_{usfe}$, making alloys intrinsically ductile.
\end{abstract}

\clearpage
\section{Introduction}

The continued cost-pressures and ever stringent environmental norms have pushed aerospace and power generation industries to improve the overall efficiency of their jet-engines and gas-turbines \cite{Baker2021}. Higher operating temperatures would lead to better thermodynamic efficiency in these propulsion and energy conversion systems \cite{Yeh2006,MURTY2019247}. The Ni-based superalloys have ruled this application domain for the past more than six decades \cite{Reed2006}. The melting point of the base metal Ni (T$_m$=1455$\degree$C) limits the highest operating temperature of these Ni-based superalloys \cite{Baker2021,Senkov2018}. Alloys based on refractory metals show great potential in this domain due to their high melting-points and their ability to maintain mechanical properties at temperatures higher than the current Ni-based superalloys \cite{Senkov2018}. Refractory alloys are based on Group IV, V, and VI metals \cite{Shaikh2020m,Miracle2019,MURTY201913}. These elements have very high melting-points (T$_m<<$1800$\degree$C) and BCC crystal structure which limits their ductility at ambient temperatures.

The lack of deformability at lower temperatures makes refractory alloys difficult to manufacture, which creates a bottleneck in developing them for various applications. Mo has been ductilized by adding 25at.\% of Ti/Zr/Hf indicating that alloying with HCP metals should ductilize Mo by reducing the Rice-Thompson parameter\cite{Geller2005}. Addition of up to 25at.\% of  Ta/Re to W has been shown to ductilize it by reducing the overall $\gamma_{usfe}$ of the alloy\cite{Yang2018}. Re addition in 12.5at.\% or 25at.\% to W has been shown to decrease the shear resistance of \{110\}\textless111\textgreater\ and \{112\}\textless111\textgreater\ slip system which in turn improves the ductility of W \cite{Li2019}. Reducing the valence electron concentration (VEC) \cite{Shaikh2020m,Sheikh2016} and reducing the lattice distortion in Nb-based equiatomic BCC alloys has been shown to decrease the Peierls stress which in turn makes alloy deformable\cite{Zhao2019}. These are few of the refractory alloy ductilizing studies reported in literature. There have been conflicting suggestions about alloying additions to make W deformable. Qian et al. \cite{Qian2018} suggests Re addition to W leads to decreased generalized stacking fault energy ($\gamma_{gsfe}$) and increased ductility, whereas Ta addition has opposite effect. However, in Qian's work the chemistry of the supercell was W$_{47}$X (X=Ta or Re), which may not be a true representation of a concenterated alloy. On the contrary, Sheikh et al. \cite{Sheikh2016} suggests alloying with Ti, Zr, Hf, and Re (HCP elements) should ductilize the refractory alloys as it reduces the valence electron concentration (VEC). Similarly, Yang et al.\cite{Yang2018} suggests alloying W with Ta or Re should ductilize it by reducing the unstable stacking-fault energy ($\gamma_{usfe}$). None of the previous strategies discuss about the thermodynamic stability of these alloys. This calls for a comprehensive study on the role of various alloying elements on the deformability and thermodynamic stability of refractory alloys.

A general theme in the existing refractory alloy ductilizing studies has been to add low stacking-fault energy (SFE, $\gamma_{sfe}$) elements to reduce the overall unstable-SFE ($\gamma_{usfe}$) of the alloy, suggesting that the Rule-of-Mixtures (ROM) dictates the overall $\gamma_{usfe}$ of the alloy. Most of the earlier studies selected alloying elements to get maximum decrease in the overall $\gamma_{usfe}$ of the alloy, without considering the thermodynamic stability of the alloy. The $\gamma_{usfe}$ of NiFe \cite{Zhao2019a,Zaddach2013} and CoCrFeNi \cite{BeyramaliKivy2017,Liu2018} calculated using first-principles density functional theory (DFT) matches with their respective experimental value. This shows that the $\gamma_{sfe}$ can be accurately predicted using DFT. There is a need of a comprehensive study on the effect of alloying elements on the deformability and thermodynamic stability of refractory BCC alloys. Present work addresses these issues using first-principles density functional theory (DFT) simulations. The study gives a strategy to select alloying elements which give maximum ductilizing effect along with thermodynamic stability of the final alloy.

\section{Results and Discussion}
To better understand the factors affecting the intrinsic ductility in concentrated BCC alloys, refractory equiatomic binary alloys made from Group-IV (Ti, Zr, Hf), Group-V(V, Nb, Ta), Group-VI(Mo, W), and Group-VII(Re) elements are studied. Atomistic modeling based on DFT provides an accurate description of configurational and chemical space. Based on the binary phase diagrams, not all refractory element combination would be stable as a solid solution in BCC crystal structure. As a first step we take only those combinations which are likely to form a stable BCC solid solution. We show that the enthalpy of formation can be a good estimate of the stability of solid solutions (Section \ref{section:Ef}). Here we report for the first time the influence of enthalpy of formation on intrinsic ductility (D) of alloys. The D is a ratio of unstable stacking-fault energy ($\gamma_{usfe}$) and surface energy ($\gamma_{s}$), which has been widely used to compare the intrinsic ductility of refractory alloys \cite{Yang2018,Hu2021c,Senkov2021d,Mak2021}. However, it needs a reliable calculation of the $\gamma_{usfe}$ and $\gamma_{s}$.

The $\gamma_{usfe}$ calculation of solid solutions using DFT, is bound to have errors due to the change in local chemistry while the interface is being sheared and due to the stoichiometry of the shearing interface. The magnitude of this error could alter conclusions, hence an estimate of this error is important while making comparison across alloys. Here we have estimated the maximum error that gets introduced, and discuss ways to minimize it, making our conclusions reliable. The $\gamma_{usfe}$ of an alloy is strongly influenced by its pure metal constituents. To better understand how pure metal's $\gamma_{usfe}$ value affect the $\gamma_{usfe}$ of the alloy, the $\gamma_{usfe}$ of similar slip systems should be taken into consideration. In present study some of the pure metals have HCP crystal structure and the alloys are considered in BCC structure. Therefore, for both the crystal structures, the $\gamma_{usfe}$ of equivalent slip system should be considered.

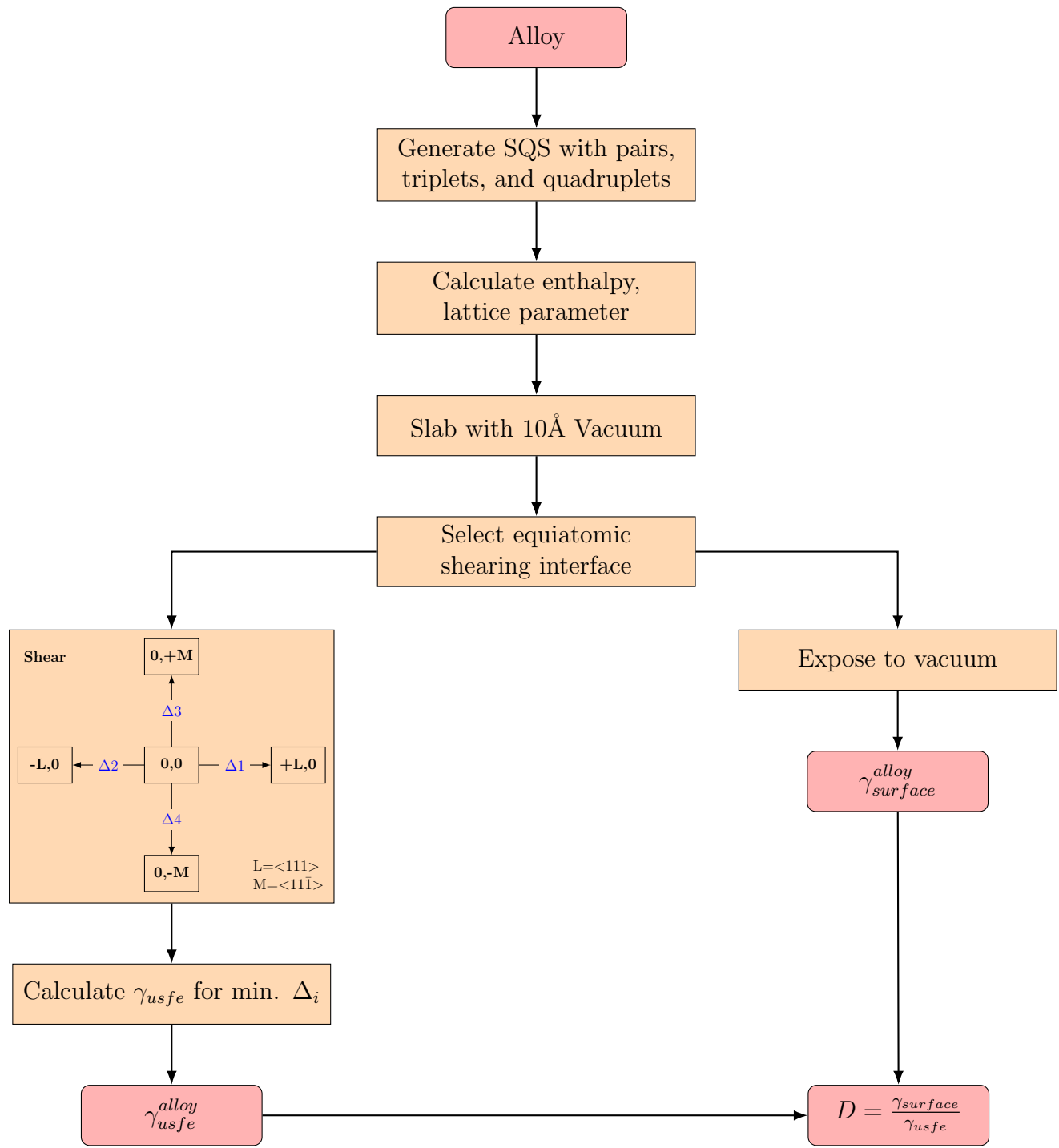
\begin{figure}[!h]
\centering
    \begin{tikzpicture}[node distance=1cm]
    \tikzstyle{startstop} = [rectangle, rounded corners, minimum width=3cm, minimum height=1cm,text centered, draw=black, fill=red!30]
    \tikzstyle{process} = [rectangle, minimum width=3cm, minimum height=1cm, text centered, draw=black, fill=orange!30]
    \tikzstyle{decision} = [diamond, aspect=3, text centered, draw=black, fill=green!30]
    \tikzstyle{arrow} = [thick,-Latex]
    
    \node (alloy) [startstop] {Alloy};
    \node (bSQS) [rectangle, text width=5cm, minimum height=1.2cm, text centered, draw=black, fill=orange!30, below =of alloy] {Generate SQS with pairs, triplets, and quadruplets};
    \node (latticeParam) [rectangle, text width=5cm, minimum height=1.2cm, text centered, draw=black, fill=orange!30, below =of bSQS] {Calculate enthalpy, lattice parameter};
    \node (supercell) [process, text width=5cm, below =of latticeParam] {Slab with 10Å Vacuum};
    \node (eqPlane) [process, below =of supercell, text width=5cm] {Select equiatomic shearing interface};
    \node (12A) [process, below left =of eqPlane] {
    \begin{tikzpicture}[scale=0.6,every node/.style={scale=0.6}]
        \tikzstyle{hello} = [draw,minimum width=1.5cm,minimum height=1cm]
        \tikzstyle{arrowtext} = [anchor=center,minimum width=0.5cm, minimum height=0.5cm,fill=orange!30]
        \node (00) at (0,0) [hello] {\textbf{0,0}};
        \node (+0) at (3.5,0) [hello] {\textbf{+L,0}};
        \node (-0) at (-3.5,0) [hello] {\textbf{-L,0}};
        \node (0+) at (0,3) [hello] {\textbf{0,+M}};
        \node (0-) at (0,-3) [hello] {\textbf{0,-M}};
        \node (-+) at (-3.5,+3) [minimum width=1.5cm,minimum height=1cm] {\textbf{Shear}};
        \node (-+) at (+3.5,-3) [align=right,text width=1cm,minimum height=1cm] {L=\textless111\textgreater\\M=\textless11$\bar{1}$\textgreater};
        \draw[-latex] (00) -- node[arrowtext]{\textcolor{blue}{$\Delta1$}} (+0);
        \draw[-latex] (00) -- node[arrowtext]{\textcolor{blue}{$\Delta2$}} (-0);
        \draw[-latex] (00) -- node[arrowtext]{\textcolor{blue}{$\Delta3$}} (0+);
        \draw[-latex] (00) -- node[arrowtext]{\textcolor{blue}{$\Delta4$}} (0-);
    \end{tikzpicture}};
    \node (gamma12) [process, below =of 12A, text width=5cm] {Calculate $\gamma_{usfe}$ for min. $\Delta_i$};
    \node (finalGamma) [startstop, below =of gamma12] {$\gamma_{usfe}^{alloy}$};
    \node (expose) [process, text width=5cm, below right =of eqPlane] {Expose to vacuum};
    \node (surfacE) [startstop, below =of expose] {$\gamma_{surface}^{alloy}$};
    \node (ductilityParam) [startstop, right =of finalGamma, xshift=8.05cm] {$D=\frac{\gamma_{surface}}{\gamma_{usfe}}$};
    
    \draw[arrow] (alloy.south) -- (bSQS.north);
    \draw[arrow] (bSQS.south) -- (latticeParam.north);
    \draw[arrow] (latticeParam.south) -- (supercell.north);
    \draw[arrow]  (supercell.south) -- (eqPlane.north);
    \draw[arrow]  (eqPlane.west) -| (12A.north);
    \draw[arrow]  (eqPlane.east) -| (expose);
    \draw[arrow]  (expose) -- (surfacE);
    \draw[arrow] (12A.south) -- (gamma12);
    \draw[arrow] (gamma12.south) -- (finalGamma);
    \draw[arrow] (finalGamma.east) -- (ductilityParam.west);
    \draw[arrow] (surfacE) -- (ductilityParam);
    \end{tikzpicture}
\caption{Workflow.}
\label{fig:flowchart}
\end{figure}

To address these shortcomings which we found in the literature, a DFT-based workflow is developed to calculate lattice parameter (a,c), enthalpy of formation ($\Delta E_f$), unstable stacking-fault energy ($\gamma_{usfe}$), and surface energy ($\gamma_s$) (Figure \ref{fig:flowchart}). The workflow is elaborated further in the subsequent sections. The results are shown in the form of heat maps in Figure \ref{fig:heat maps}. Additional information can be found in the supplementary file.

\subsection{Enthalpy of formation ($\Delta E_f$)}
\label{section:Ef}
The stability of an alloy (solid solution) is dictated by the Gibbs energy of its constituent phases as,
\begin{align*}
    \Delta G &= \Delta H - T\Delta S \\
    &= \Delta E_f + {\Delta(PV)} - T\{\Delta S_{config} + \Delta S_{vib} + \Delta S_{mag} + \Delta S_{elec}\}
\end{align*}
where $\Delta E_f$ is the enthalpy of formation at 0K, $\Delta(PV)$ is the pressure-volume term, $\Delta S_{config}$ is the configurational entropy, $\Delta S_{vib}$ is the vibrational entropy, $\Delta S_{mag}$ is the magnetic entropy, $\Delta S_{elec}$ is the electronic entropy. The $\Delta(PV)$ term can be ignored \cite{VandeWalle2002}. Therefore the equation becomes,
\begin{equation*}
\label{eq:DeltaG}
    \Delta G = \Delta E_f - T\Delta S_{config} - T\{\Delta S_{vib} + \Delta S_{mag} + \Delta S_{elec}\}
\end{equation*}
For an equiatomic binary alloy AB, $\Delta E_f$ is calculated with the relation \cite{Shin2007},
\begin{equation*}
   \Delta E_{f}^{AB}=E_{AB}-0.5*(E_A+E_B)
\end{equation*}
where E$_{AB}$ is the energy per atom of alloy AB in BCC symmetry, E$_A$ and E$_B$, are the energy of constituents A and B in their most stable phase. All values are in the units of eV/Atom. The 9 alloying elements form a total of 36 equiatomic binary alloys. Out of these 36 alloys, 6 systems containing both the constituents being HCP, are not considered in the present study. Figure \ref{fig:heatmapEnthalpy} shows the enthalpy of formation ($\Delta E_f$, eV/Atom) values in the form of heat map. We have not considered VRe \cite{Villars2016:sm_isp_c_0902814}, TaRe \cite{Villars2016:sm_isp_c_0100260}, HfW \cite{Villars2016:sm_isp_c_0907545}, ZrW \cite{Villars2016:sm_isp_c_0904967}, VHf \cite{Villars2016:sm_isp_c_0100184}, VZr \cite{Villars2016:sm_isp_c_0905433}, having $\Delta E_f$, -0.2456, -0.1942, 0.1114, 0.1477, 0.1631, 0.1710 meV/Atom respectively, for subsequent analysis as they can either form intermetallic (VRe, TaRe) due to very negative $\Delta E_f$ or they can segregate (HfW, ZrW, VHf, VZr) due to very positive $\Delta E_f$. The $\Delta E_f$ values of alloy considered ranges from -0.1232eV/Atom (TaMo) to 0.0994eV/Atom (WRe).

The $\Delta E_f$ value of 0.0994meV/Atom appear very large for a solid solution to be stable. For an equiatomic binary alloy, the T$\Delta S_{config}$ term has value of about 18 meV/Atom at 300K. Since WRe forms a stable solid solution \cite{Villars2016:sm_isp_c_0904355}, we expect that the other entropic contributions would be sufficiently large ($\approx$80meV/Atom) to overcome the 0.0994eV/Atom $\Delta E_f$ value. Although we take WRe as an extreme case of having the most positive $\Delta E_f$ (among the presently considered alloys), depending upon the $T\Delta S_{config}$ and $T\Delta S_{vib}$ values, not all alloys may be stable.

\subsection{Unstable stacking-fault energy ($\gamma_{usfe}$)}
\label{section:usfe}
    
Within DFT, the $\gamma_{usfe}$ is calculated using the relation given below,
\begin{equation*}
    \gamma_{usfe} = \frac{E_{faulted}-E_{pristine}}{(Area\ of\ Plane)}
\end{equation*}
where the E$_{faulted}$ is the energy of supercell having a stacking fault and the E$_{pristine}$ is the energy of pristine supercell. The \{110\}\textless111\textgreater\ slip system of BCC metals/alloys has the lowest energy barrier for activation. Therefore, we have chosen the same slip system in present anaysis.

\subsubsection{Slip system}
\label{section:slipSystem}
\begin{figure}
	\begin{subfigure}[t]{0.5\columnwidth}
	\centering
		\includegraphics[width=\columnwidth]{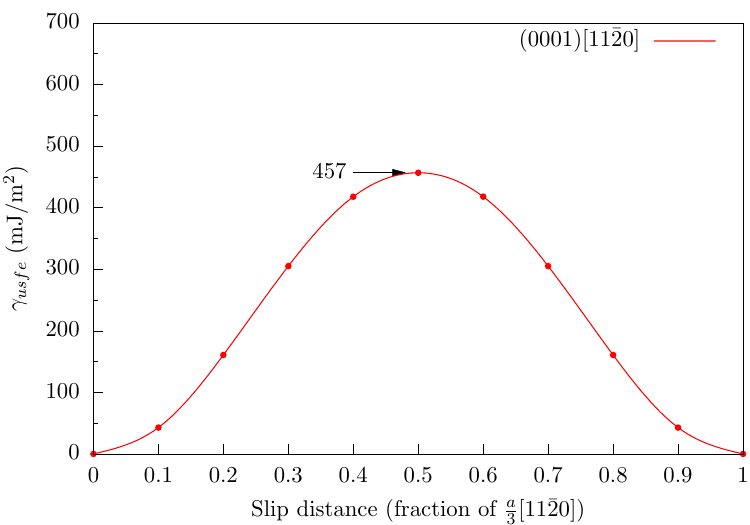}
		\caption{}
		\label{fig:hcp usfe curve}
	\end{subfigure}
	\begin{subfigure}[t]{0.5\columnwidth}
	\centering
		\resizebox{0.97\columnwidth}{!}{
			\begin{tikzpicture}
				\draw [color=blue, fill=blue] (0, 0) circle (0.1);
				\draw [color=blue, fill=blue] (3,0) circle (0.1);
				\draw [color=blue, fill=blue] (-1.5,{3*sqrt(3)/2}) circle (0.1);
				\draw [color=blue, fill=blue] (1.5,{3*sqrt(3)/2}) circle (0.1);
				
				\draw[blue, very thin] (0,0) -- (3,0);
				\draw[blue, very thin] (0,0) -- (-1.5,{3*sqrt(3)/2});
				\draw[blue, very thin] (1.5,{3*sqrt(3)/2}) -- (3,0);
				
				\draw[black,-latex,thick] (-1.5,{3*sqrt(3)/2}) -- (3,0);
				\draw[red,-latex,thick] (-1.5,{3*sqrt(3)/2}) -- (1.5,{3*sqrt(3)/2});
				
				\node at (3,-0.5) {\textcolor{black}{\textless1$\bar{1}$00\textgreater}};
				\node at (2,{0.5+3*sqrt(3)/2}) {\textcolor{red}{\textless11$\bar{2}$0\textgreater}};
		\end{tikzpicture}}
		\caption{}
		\label{fig:hcp usfe projection}
	\end{subfigure}
	\begin{subfigure}[t]{0.5\columnwidth}
	\centering
		\includegraphics[width=\columnwidth]{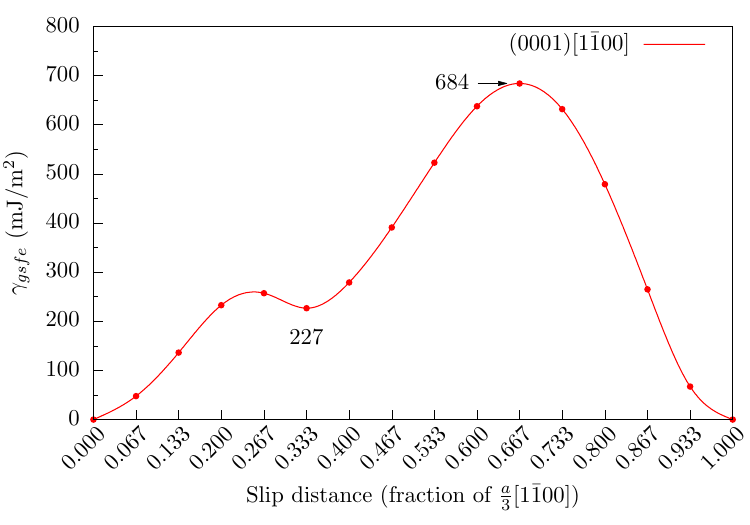}
		\caption{}
		\label{fig:hcp sfe curve}
	\end{subfigure}
	\begin{subfigure}[t]{0.5\columnwidth}
	\centering
		\resizebox{0.97\columnwidth}{!}{
			\begin{tikzpicture}
				\draw [color=blue, fill=blue] (0, 0) circle (0.1);
				\draw [color=blue, fill=blue] (3,0) circle (0.1);
				\draw [color=blue, fill=blue] (-1.5,{3*sqrt(3)/2}) circle (0.1);
				\draw [color=blue, fill=blue] (1.5,{3*sqrt(3)/2}) circle (0.1);
				
				\draw[blue, very thin] (0,0) -- (3,0);
				\draw[blue, very thin] (0,0) -- (-1.5,{3*sqrt(3)/2});
				\draw[blue, very thin] (1.5,{3*sqrt(3)/2}) -- (3,0);
				
				\draw[red,-latex,thick] (-1.5,{3*sqrt(3)/2}) -- (3,0);
				\draw[black,-latex,thick] (-1.5,{3*sqrt(3)/2}) -- (1.5,{3*sqrt(3)/2});
				
				\node at (3,-0.5) {\textcolor{red}{\textless1$\bar{1}$00\textgreater}};
				\node at (2,{0.5+3*sqrt(3)/2}) {\textcolor{black}{\textless11$\bar{2}$0\textgreater}};
		\end{tikzpicture}}
		\caption{}
		\label{fig:hcp sfe projection}
	\end{subfigure}
    \begin{subfigure}[t]{0.5\columnwidth}
    \centering
		\includegraphics[width=\columnwidth]{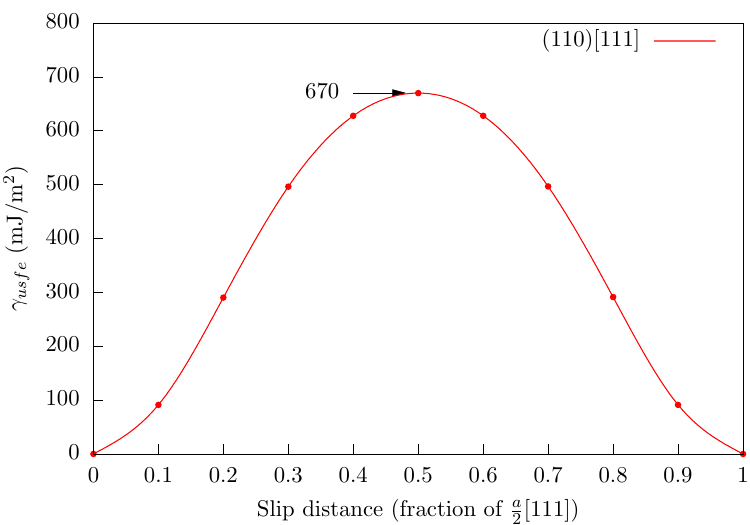}
		\caption{}
		\label{fig:bcc usfe curve}
	\end{subfigure}
	\begin{subfigure}[t]{0.5\columnwidth}
	\centering
 		\resizebox{\columnwidth}{!}{
			\begin{tikzpicture}

				
				
				\draw [color=blue, fill=blue] (0, 0) circle (0.1);
				\draw [color=blue, fill=blue] (3, 0) circle (0.1);
				\draw [color=blue, fill=blue] ({-3*0.34},{3*0.94}) circle (0.1);
				\draw [color=blue, fill=blue] ({3-3*0.34},{3*0.94}) circle (0.1);
				
				\draw[blue, very thin] (3,0) -- ({3-3*0.34},{3*0.94}) -- ({-3*0.34},{3*0.94});
				\draw[red, -latex, thick] (0,0) -- (3,0);
				\draw[black, -latex, thick] (0,0) -- ({-3*0.34},{3*0.94});
				
				\node at (3,-0.5) {\textcolor{red}{\textless111\textgreater}};
				\node at ({-3.25*0.34},{3.35*0.94}) {\textcolor{black}{\textless11$\bar{1}$\textgreater}};

		\end{tikzpicture}
 		}
		\caption{}
		\label{fig:bcc usfe projection}
	\end{subfigure}

\caption{BCC and HCP slip system comparison. (a)Zr (0001)[$11\bar{2}0$] slip curve, (b)HCP (0001)[$11\bar{2}0$] slip system projection along [0001] direction, (c)Zr (0001)[$1\bar{1}00$] slip curve, (d)HCP (0001)[$1\bar{1}00$] slip system projection along [0001] direction, (e)Nb (110)[111] slip curve, (f)BCC (110)[111] slip system projection along [110] direction. (Out of plane atoms are not shown)}
\label{fig:stacking fault curves and projections}
\end{figure}
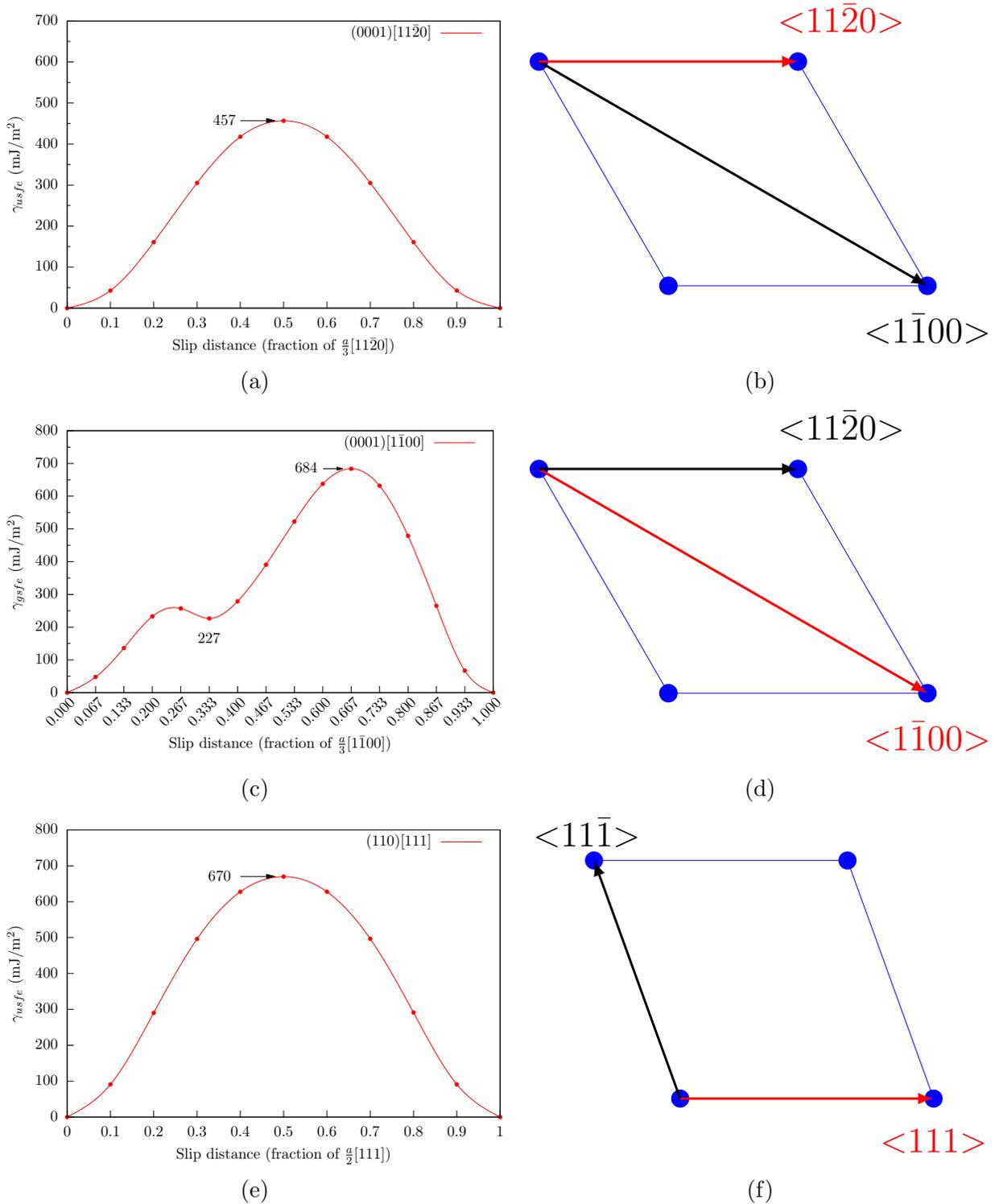

The $\gamma_{usfe}$ of an alloy strongly depends on the $\gamma_{usfe}$ of its constituent pure metals. In order to understand how the $\gamma_{usfe}$ of pure elements affect the $\gamma_{usfe}$ of an alloy, comparison of similar similar slip systems should be made. Since some of the pure metal constituents of alloy have HCP crystal structure, the \{110\}\textless111\textgreater\ slip system of BCC should be compared with an equivalent slip system in HCP crystal structure. The \{0001\} slip plane in HCP crystal structure have similar close-packed arrangement of atoms as that in \{110\} slip plane of BCC crystal structure. The \{0001\} slip plane SFE curves of Zr are shown in Figure \ref{fig:hcp usfe curve} for \textless11$\bar{2}$0\textgreater\ and in Figure \ref{fig:hcp sfe curve} for \textless10$\bar{1}$0\textgreater\ slip direction. Figure \ref{fig:bcc usfe curve} shows the SFE curve of Nb for \{110\}\textless111\textgreater\ slip system. From Figures \ref{fig:hcp usfe projection} and \ref{fig:bcc usfe projection}, it is observed that the atomic arrangements in BCC-\{110\} and HCP-\{0001\} slip planes is similar. Therefore, from Figure \ref{fig:stacking fault curves and projections} it is clear that the \{110\}\textless111\textgreater\ $\gamma_{usfe}$ of BCC crystal structure should be compared with the \{0001\}\textless11$\bar{2}$0\textgreater\ $\gamma_{usfe}$ of HCP crystal structure.

\subsubsection{Shearing interface stoichiometry}
\label{section:shearingInterfaceChemistry}
Special quasirandom structures (SQS) have been extensively used to predict the SFE of alloys due to their simplicity in capturing the inherent chemical disorder present in the alloys \cite{Zunger1990}. In present work, the SQS supercell with in-plane dimensions equal to 4 and 3 times of the first nearest neighbor distance of the BCC symmetry are generated (Figure S1), considering the pair, triplet, and quadruplet with cut-off distance equal to the regular BCC unit cell lattice parameter (2$^{nd}$ nearest neighbor distance). The supercell have 10 planes of (110) type giving 9 shearing interfaces as shown in Figure S1. A vacuum of 10Å is added to the supercell to prevent the interactions due to the periodic boundary condition. To calculate the $\gamma_{usfe}$, the atoms in top two planes and bottom two planes are fixed in all directions, whereas, the remaining atoms from the middle six planes are fixed only in slip direction. There are four [111]-type slip directions available for every shearing interface, giving 4 different $\gamma_{usfe}$ values. Therefore, the same supercell can gives 36 different $\gamma_{usfe}$ values.

\begin{figure}
\centering
        \includegraphics[width=0.5\columnwidth]{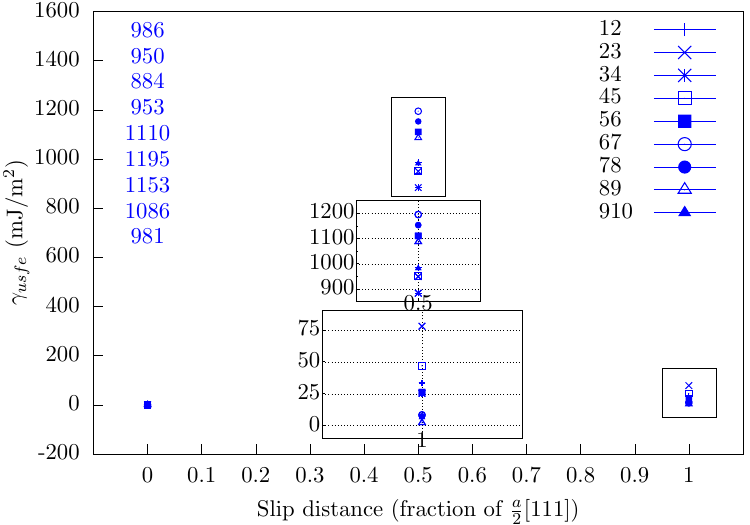}
\label{fig: shearing interfaces curves}
\caption{Comparision between $\gamma_{usfe}$\ values calculated for various shearing interfaces from same supercell of WRe. Legend shows the stiochiometry of the shearing interface. Numbers on left shows the $\gamma_{usfe}$ value. Upper inset shows the range of $\gamma_{usfe}$. Lower inset shows the energy difference between before and after one complete slip by $\bar{b}$=$\frac{a}{2}[111]$.}
\label{fig:WRe shearing interfaces}
\end{figure}

Using the above methodology, we calculated the $\gamma_{usfe}$ for (110)[111] slip system of WRe alloy (Figure \ref{fig:WRe shearing interfaces}). For any shearing interface the energy change in the pristine and slipped supercell is due to the change in the local bonding environment. Out of the 4 possible $\gamma_{usfe}$, we choose the one with the minimum difference in energy of pristine and sheared slab as shown in Figure S2 and S3. Therefore there are 9 $\gamma_{usfe}$ values shown in in Figure \ref{fig:WRe shearing interfaces} instead of 36. As observed from Figure \ref{fig:WRe shearing interfaces}, the $\gamma_{usfe}$ varies from 906 to 1199 mJ/m$^2$. Similarly the energy difference between pristine and sheared supercell varies from 0.5 to 109 mJ/m$^2$ (lower inset in Figure \ref{fig:WRe shearing interfaces}). There are two shearing interfaces in Figure \ref{fig:WRe shearing interfaces} having the same stoichiometry as that of the overall supercell chemistry (WRe). These two shearing interfaces give 1110mJ/m$^2$ and 1086mJ/m$^2$ $\gamma_{usfe}$ value. It indicates that the 2$^{nd}$ nearest neighbor also influences the $\gamma_{usfe}$. We have taken 1110mJ/m$^2$ as the $\gamma_{usfe}$ value of WRe in present study as this shearing interface was having the lowest energy difference of 1.6mJ/m$^2$ between the pristine and sheared supercell.

Based on the above discussion, one should consider only the equiatomic shearing interface. Since the supercell has shearing interfaces that do not have equiatomic stoichiometry, it is assumed that all possible values of $\gamma_{usfe}$ can exist. The strategy in the literature have been to report an average $\gamma_{usfe}$ obtained from all shearing interfaces. Such approach does not assure a specific value of error that gets introduced. From WRe results (Figure \ref{fig:WRe shearing interfaces}), it clear that stoichiometry strongly affect the $\gamma_{usfe}$. That is because of nearest neighbor bond in the shearing interface. Since the intent is to calculate the $\gamma_{usfe}$ for an equiatomic alloy, we choose the value corresponding to equiatomic shearing interface. Among the two interfaces that have same stoichiometry in Figure \ref{fig:WRe shearing interfaces}, there is a difference in $\gamma_{usfe}$ value due to different set of second nearest neighbors. For the rest of the alloys, the obvious choice of $\gamma_{usfe}$ will be from the shearing interface having the same stoichiometry as that of the supercell. We found that the maximum energy difference between the pristine and the sheared supercell is 60mJ/m$^2$. This error is alloy dependent, whereas the lowest error is 0mJ/m$^2$ in NbTa. Such error estimation which could affect the interpretation have not been discussed in the literature.

\subsubsection{On the origin of $\gamma_{usfe}$ of binary alloys}
\begin{figure}
	\begin{subfigure}{0.50\columnwidth}
        \includegraphics[width=\columnwidth]{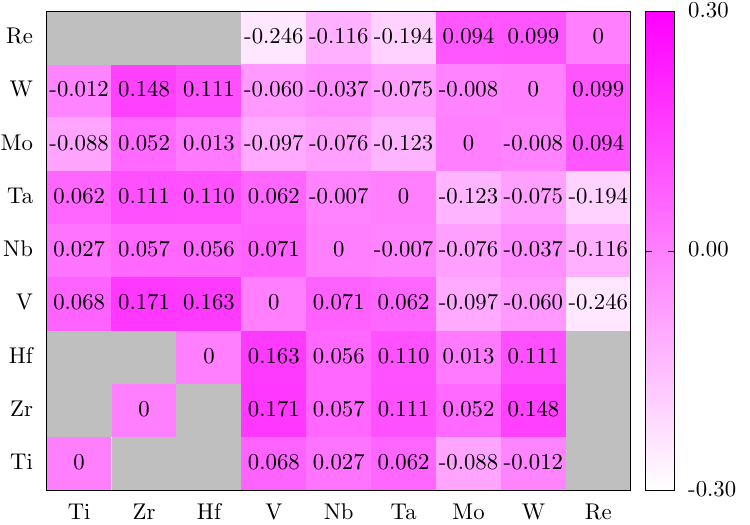}
		\caption{Enthalpy of formation (eV/Atom)}
		\label{fig:heatmapEnthalpy}
	\end{subfigure}
	\begin{subfigure}{0.50\columnwidth}
        \includegraphics[width=\columnwidth]{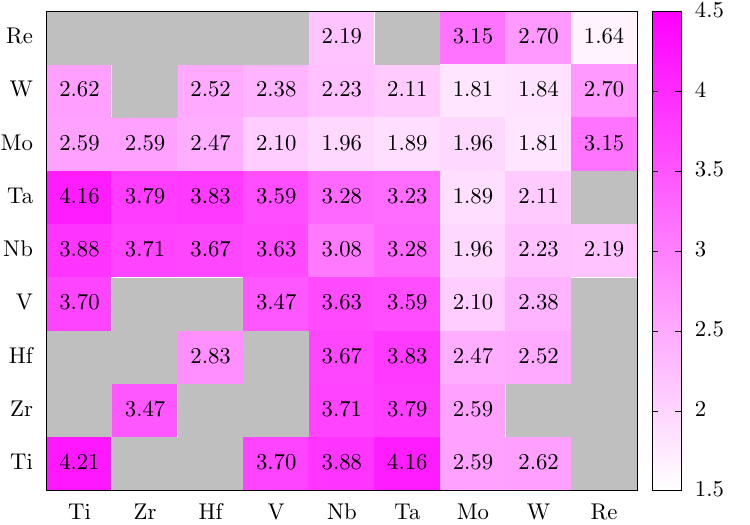}
		\caption{D-parameter}
		\label{fig:heatmapD}
	\end{subfigure}
	\begin{subfigure}{0.50\columnwidth}
        \includegraphics[width=\columnwidth]{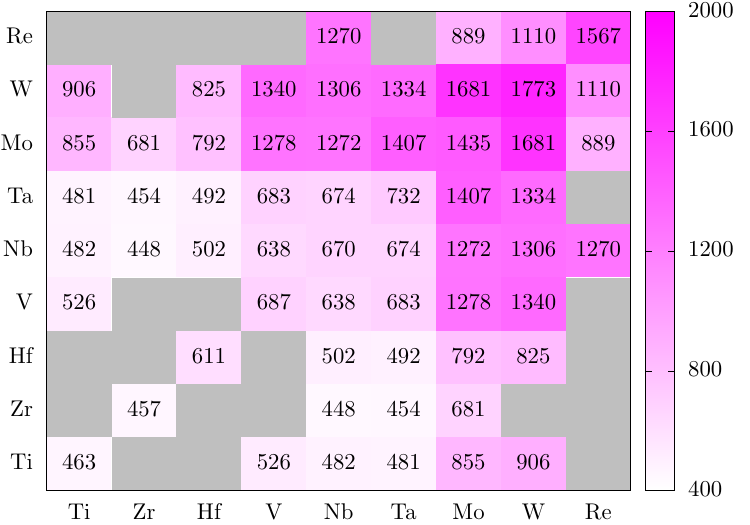}
		\caption{$\gamma_{usfe}$\ (mJ/m$^2$)}
		\label{fig:heatmapUSFE}
	\end{subfigure}
	\begin{subfigure}{0.50\columnwidth}
        \includegraphics[width=\columnwidth]{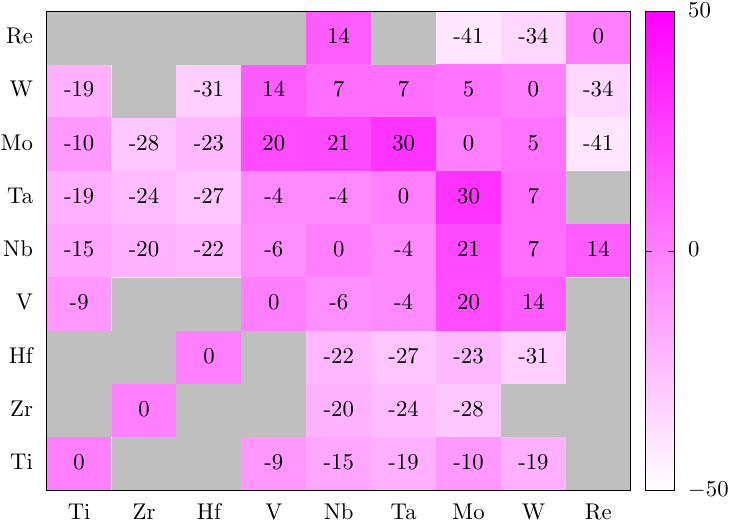}
		\caption{Change in $\gamma_{usfe}$\ (\%)}
		\label{fig:heatmap100DeltaUSFE}
	\end{subfigure}
	\begin{subfigure}{0.50\columnwidth}
        \includegraphics[width=\columnwidth]{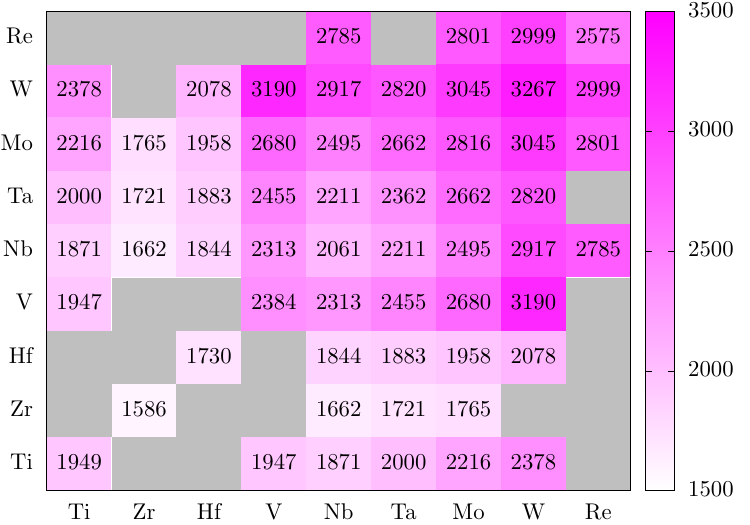}
		\caption{$\gamma_{s}$\ (mJ/m$^2$)}
		\label{fig:heatmapSurfacE}
	\end{subfigure}
	\begin{subfigure}{0.50\columnwidth}
        \includegraphics[width=\columnwidth]{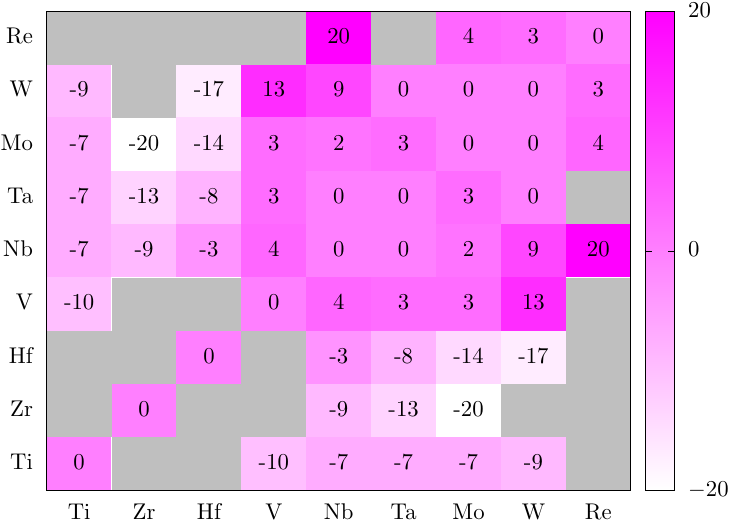}
		\caption{Change in $\gamma_{s}$\ (\%)}
		\label{fig:heatmap100DeltaSurfacE}
	\end{subfigure}
	\caption{Heat maps. Alloys containing both HCP constituents are not considered.}
	\label{fig:heat maps}
\end{figure}

Figure \ref{fig:heatmapUSFE} shows the $\gamma_{usfe}$ calculated using First-principles DFT method, in the form of heat map for pure elements and the alloys under study. The HCP elements have the lowest $\gamma_{usfe}$ except Re. For pure metals, the overall trend is of increasing $\gamma_{usfe}$ as we move right in the periodic table from Group-IV to Group-VII. W (1773mJ/m$^3$) has the highest $\gamma_{usfe}$ whereas Zr (457mJ/m$^3$) has the lowest $\gamma_{usfe}$. For alloys, the DFT-calculated $\gamma_{usfe}$ ranges from 454mJ/m$^2$ (TaZr) to 1681mJ/m$^2$ (MoW). Figure \ref{fig:heatmap100DeltaUSFE} shows the change in $\gamma_{usfe}$ of the alloys from their composition averaged value ($\Delta\gamma_{usfe}$) as calculated below,
\begin{equation*}
    \Delta\gamma_{usfe} = \frac{\gamma^{DFT}_{usfe}-\gamma^{ROM}_{usfe}}{\gamma^{ROM}_{usfe}}\times100
\end{equation*}
The $\Delta\gamma_{usfe}$ ranges from -41\% (MoRe) to +20\% (MoV). As observed from Figures \ref{fig:heatmapUSFE} and Figure \ref{fig:heatmap100DeltaUSFE}, the ROM overestimates the $\gamma_{usfe}$ for a number of alloys. For example, the DFT-calculated $\gamma_{usfe}$ for WRe and MoRe is 1110mJ/m$^2$ and 889mJ/m$^2$ respectively. However the -34\% (WRe) and -41\% (MoRe) $\Delta\gamma_{usfe}$ indicates that ROM overestimates the $\gamma_{usfe}$ by a large margin and is not a reliable method to get the correct values. This also suggests that the $\gamma_{usfe}$ of alloys is dictated by the nature of bonds between the constituent atoms.

\begin{figure}
\begin{subfigure}{0.5\columnwidth}
    \includegraphics[width=\columnwidth]{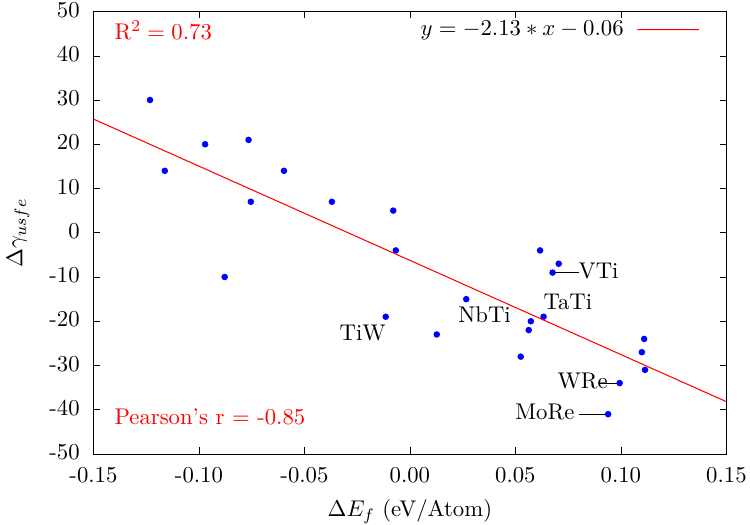}
    \caption{}
    \label{fig:correlations100deltaUSFEvsEf}
\end{subfigure}
\begin{subfigure}{0.5\columnwidth}
    \includegraphics[width=\columnwidth]{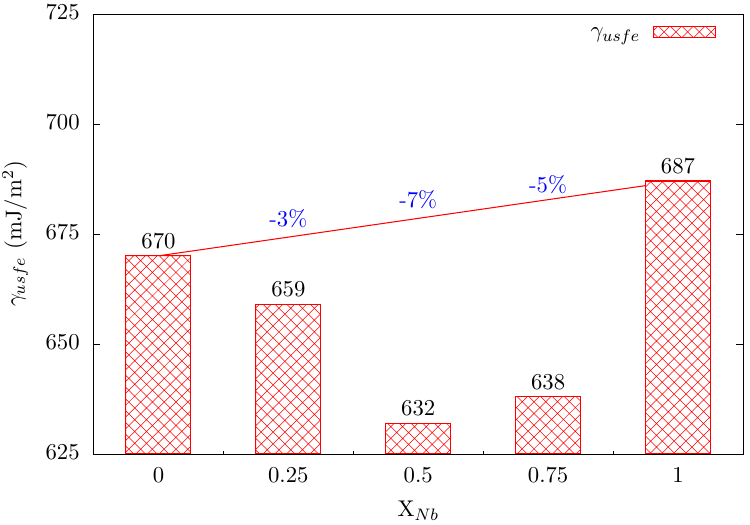}
    \caption{}
    \label{fig:minima}
\end{subfigure}
\caption{(a)Change in $\gamma_{usfe}$ as compared to the composition averaged value vs $\Delta E_f$. (b)(110)[111] $\gamma_{usfe}$  of Nb-V alloys. Numbers in blue show the change in $\gamma_{usfe}$ from the composition averaged value.}
\label{fig:correlations}
\end{figure}

The positive and negative $\Delta\gamma_{usfe}$ can be due to the repulsive or attractive interaction between the constituent atoms. One of the parameter to assess the nature of interaction is enthalpy of formation ($\Delta E_f$). The $\Delta E_f$ of equiatomic alloy captures the attractive (negative $\Delta E_f$) or repulsive (positive $\Delta E_f$) interaction between atoms even if the crystal structure of metals that form the alloy are not the same. For example, W and Re have BCC and HCP crystal structure in their pure state, respectively. The $\Delta E_f$ of W25-Re75 alloy in BCC crystal structure will have contribution from change in crystal structure of Re from HCP to BCC, apart from the interaction between atoms. In non-equiatomic alloys with different crystal structure of constituents, the $\Delta E_f$ does not reflect attractive or repulsive interaction alone, but for alloy with constituent metals in the same crystal structure, $\Delta E_f$ of any alloy chemistry will reflect the nature of interaction. Figure \ref{fig:correlations100deltaUSFEvsEf} shows the $\Delta\gamma_{usfe}$ vs. $\Delta E_f$. A linear fit to the data in Figure \ref{fig:correlations100deltaUSFEvsEf} has a slope of -2.13 with Pearson's r (correlation parameter) of -0.85. This indicates a strong inverse correlation between $\Delta E_f$ with $\Delta\gamma_{usfe}$. The $\Delta E_f$ is maximum at equiatomic composition as the number of A-B bonds in an AB alloy would be the highest at equiatomic composition. Inverse correlation between $\Delta\gamma_{usfe}$ and $\Delta E_f$ suggests that maximum $\Delta\gamma_{usfe}$ (positive or negative) would occur at equiatomic composition. Figure \ref{fig:minima} shows the $\gamma_{usfe}$ values for NbxV(1-x) (x $\in$\ [0,0.25,0.75,1]) alloy. The largest $\Delta\gamma_{usfe}$ occurs at equiatomic concentration (Figure \ref{fig:minima}). The present analysis shows many possible alloy combinations which can lead to reduced $\gamma_{usfe}$ in refractory binary alloys from their average values.

\subsection{Surface energy ($\gamma_s$)}
\label{section:surfacE}
 Similar to $\gamma_{usfe}$ calculation, the surface energy ($\gamma_s$) is calculated for the interface having equiatomic stoichiometry/formula composition. The equiatomic shearing interface of the supercell is exposed to vacuum of 10Å. The $\gamma_{s}$ is calculated using the relation given below,
\begin{equation*}
    \gamma_s = \frac{E_{with-vacuum}-E_{no-vacuum}}{2\times(Area\ of\ Plane)}    
\end{equation*}
where the E$_{with-vacuum}$ is the energy per atom of the supercell with vacuum and E$_{no-vacuum}$ is the energy per atom of the supercell without vacuum. Figure \ref{fig:heatmapSurfacE} shows the (110) $\gamma_{s}$ calculated using First-principles DFT method, in the form of heat map for pure elements and the alloys under study. For Ti, Zr, Hf, and Re, the $\gamma_{s}$ is reported for (0001) plane. The $\gamma_s$ ranges from 3190mJ/m$^2$ (VW) to 1662mJ/m$^2$ (NbZr). Figure \ref{fig:heatmap100DeltaSurfacE} shows the percentage change in $\gamma_{s}$ values of the alloys from their composition averaged values ($\Delta\gamma_s$, calculated similar to $\Delta\gamma_{usfe}$). The $\Delta\gamma_{s}$ ranges from -20\% (MoZr) to 20\% (NbRe). Figure \ref{fig:correlations100deltaSurfacEvsEf} shows the $\Delta\gamma_s$ vs. the E$_f$. A linear fit to the data in Figure \ref{fig:correlations100deltaDvsEf} has a slope of -0.62 with Pearson's r (correlation parameter) of -0.49. This indicates that the $\Delta E_f$ does not strongly influences the $\gamma_s$.

\subsection{Ductility parameter (D)}
\label{section:D}
    
\begin{figure}
\begin{subfigure}{0.5\columnwidth}
    \includegraphics[width=\columnwidth]{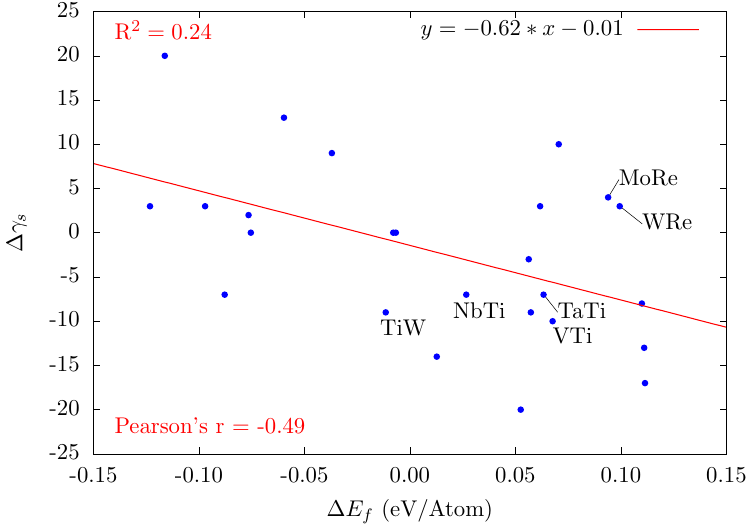}
    \caption{}
    \label{fig:correlations100deltaSurfacEvsEf}
\end{subfigure}
\begin{subfigure}{0.5\columnwidth}
\centering
    \includegraphics[width=\columnwidth]{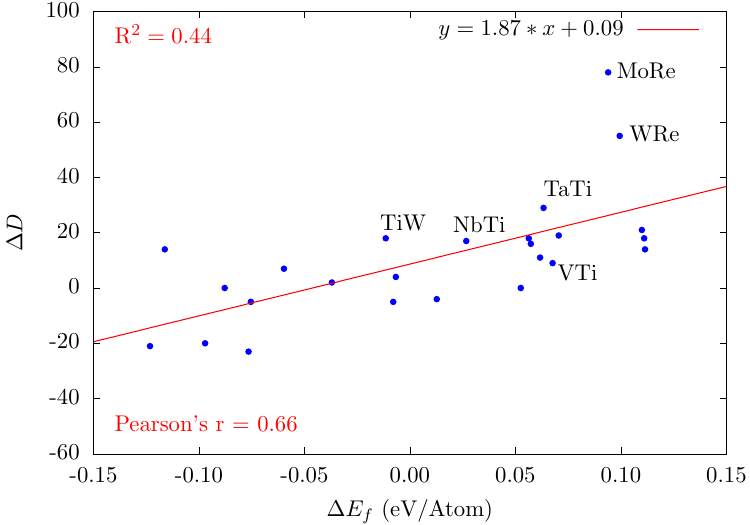}
    \caption{}
    \label{fig:correlations100deltaDvsEf}
\end{subfigure}
\caption{(a)Change in $\gamma_{s}$ as compared to the composition averaged value vs $\Delta E_f$. Enthalpy of formation does not affect the $\gamma_s$ of alloys. (b)Change in intrinsic ductility (D) as compared to the composition averaged value vs $\Delta E_f$. Positive $\Delta E_f$ helps improving the deformability of refractory binary alloys.}
\label{fig:correlations2}
\end{figure}

Ductility parameter is defined as the ratio of surface energy to unstable stacking fault energy \cite{Rice1992,Waghmare1998}.
\begin{equation*}
    \begin{aligned}
        D &= \frac{\gamma_s}{\gamma_{usfe}}\\
        D>1 \Longrightarrow \gamma_s &> \gamma_{usfe} \Longrightarrow \textbf{Intrinsically Ductile}\\
        D<1 \Longrightarrow \gamma_s &< \gamma_{usfe} \Longrightarrow \textbf{Intrinsically Brittle}
    \end{aligned}
\end{equation*}
It quantifies the competition between the energy cost of creating a new crack surface and the energy cost of nucleating new dislocations in the stress field of the crack tip. Recent studies have relied on D to assess the intrinsic ductility of refractory alloys \cite{Yang2018,Hu2021c,Senkov2021d,Mak2021}. Figure \ref{fig:heatmapD} shows the D calculated for pure metals and alloys in the form of heat map. The D ranges from 4.16 (TaTi) to 1.81 (MoW). As discussed in Section \ref{section:usfe} and \ref{section:surfacE}, the $\Delta E_f$ influences $\gamma_{usfe}$ much more strongly than $\gamma_s$. Similarly, to understand the relation between $\Delta E_f$ and percentage change in D ($\Delta D$, similar to $\Delta\gamma_{usfe}$) from its composition averaged value, we plot $\Delta D$ vs. $\Delta E_f$ in Figure \ref{fig:correlations100deltaDvsEf}. A linear fit to the data in Figure \ref{fig:correlations100deltaDvsEf} has a slope of 1.87 with Pearson's r (correlation parameter) of 0.66. The $\gamma_s$ does not show any specific relationship with E$_f$ as discussed in Section \ref{section:surfacE}. However, the $\gamma_{usfe}$ shows a strong relationship with E$_f$ as discussed in Section \ref{section:usfe}. This indicates a positive influence of $\Delta E_f$ on controlling the D of refractory binary alloys. There is a positive correlation between $\Delta D$ and $\Delta E_f$, but it is not as strong as that of $\gamma_{usfe}$, since it has contribution from $\gamma_s$ as well.

The WRe and MoRe has the largest positive $\Delta D$ as observed from Figure \ref{fig:correlations100deltaDvsEf}. Large positive $\Delta D$ of Re containing binary alloys is because of large negative $\Delta \gamma_{usfe}$ (Figure \ref{fig:correlations100deltaUSFEvsEf}) primarily due to repulsive nature of bonds between Re-W and R-Mo atoms; and large negative $\Delta \gamma_s$ (Figure \ref{fig:correlations100deltaSurfacEvsEf}). A positive $\Delta D$ is observed for alloy containing Group-IV (Ti, Zr, Hf) elements. The positive $\Delta D$ explains large ductility reported in Al/Nb/Mo-Ta-Ti-Zr-Hf-V alloy, its sub-systems \cite{Sheikh2016,Schuh2018,Juan2016,ChenPeierlsAlloy,Casillas-Trujillo2020,Soni2018} and NbTaTiZrHf \cite{Senkov2011,Senkov2015b}. This could be due to large negative $\Delta\gamma_{usfe}$, although $\Delta\gamma_s$ is not favoring $\Delta D$ to be positive. The ductility of alloys containing Ti, Zr, and Hf is attributed to the low VEC (\textless 4.5) \cite{Sheikh2016}. However, VEC criteria fails to explain the role of higher valency Re addition in improving ductility of W and Mo \cite{Li2019,Geller2005,Ren2018}. The present analysis shows that the enthalpy of formation of alloys gives a fairly reliable idea about deformability of refractory alloys apart from dictating their thermodynamic stability; which can be used as a criteria to design new alloy chemistries with desired ductility.


\section{Equations}
Not applicable.

\section{Tables}
Not applicable.

\clearpage
\section{Figures}

\section{Algorithms, Program codes and Listings}\label{algos}
Not applicable.

\section{Methods}
\label{section:methods}

\subsection{Special Quasirandom Structures, Supercell}
\label{section:workflow}
Special quasirandom structures (SQS) are used to capture chemical disorder in the alloys. SQS are generated using MCSQS code from Alloy Theoretic Automatic Toolkit (ATAT) \cite{Walle2003,VanDeWalle2013} with pair, triplet,
and quadruplet correlations with cut-off distance equal to the BCC unit cell lattice parameter (2$^{nd}$ nearest neighbor distance). The enthalpy of formation ($E_f$) and lattice parameter are calculated with 128 atoms SQS generated using a BCC unit cell repeated four times each in \textless100\textgreater\ directions. The experimental values of lattice parameters are taken from Ref.\cite{Xu2011}. The reported lattice parameters were within 1\% deviation of their experimental values. The supercells were visualized using VESTA software \cite{Momma:db5098}.

\subsection{First-principles calculations}
The first-principles density functional theory (DFT) calculations were performed using Vienna Ab-initio Simulation Package (VASP) with plane-wave basis and projector augmented wave (PAW) pseudopotentials \cite{Kresse1993,Blochl1994,Kresse1996}. For all calculations, a plane wave kinetic energy cutoff of at least 1.3 times the maximum given in the pseudopotential was used. The electronic exchange-correlation effects were calculated by Perdew-Burke-Erzernhoff generalized gradient approximation (PBE-GGA) \cite{Perdew1996,Joubert1999}. Methfessel-Paxton smearing method with 0.2 eV width was used \cite{Methfessel1989}. Structural relaxation was terminated when the forces on atoms become less than 1 meV/Å. Tetrahedron method with Blöch correction was used for energy calculation \cite{Blochl1994}. The Brillouin zone sampling was performed using Monkhorst–Pack \cite{Hu2019} scheme with automatically generated mesh with k-point spacing of less than $2\pi\times0.03$ Å$^{-1}$.


\section{Conclusion}
Here we used DFT to calculate the enthalpy of formation ($\Delta E_f$), unstable stacking-fault energy ($\gamma_{usfe}$), surface energy ($\gamma_s$), and intrinsic ductility (D) of concentrated alloys. We found that the first nearest neighbor has the strongest influence on the $\gamma_{usfe}$, hence the shearing interface with formula composition has been chosen for calculating the $\gamma_{usfe}$. The calculated $\gamma_{usfe}$ had maximum error of $\pm$30mJ/m$^2$. The E$_f$ of the equiatomic binary alloys ranges from -0.25eV/Atom to 0.17eV/Atom. The negative E$_f$ indicates attractive nature of the bonding between the constituents and vice-versa. Here we have shown that the positive $\Delta E_f$ shall lead to reduced $\gamma_{usfe}$ (compared to the composition averaged value) due to the repulsive interaction between the alloy constituents.

Our results suggests that the maximum reduction in $\gamma_{usfe}$ could be achieved for alloys having positive $\Delta E_f$. Therefore while selecting the alloying elements one should ensure the positive enthalpy of formation is well compensated by the sufficiently large entropy. Our findings provide an explanation for the addition of low valency Ti, Zr, and Hf as well as high valency Re in improving the ductility of refractory alloys. Our findings also explain the failure of empirical rules for increased ductility of refractory BCC alloys where VEC type criterion can be contradictory to experimental observations (e.g. Re addition in W). Taking the effect of enthalpy of formation on the deformability of concentrated alloys is likely to open new directions in the design of refractory alloys for high temperature applications.

\section{Declarations}
\subsection{Funding}
We acknowledge the use of the computing resources at High Performance Computing Environment (HPCE), IIT Madras. This work was supported by Ministry of Education (formerly known as Ministry of Human Resource Development), Government of India (grant numbers: SB20210844MMMHRD008277, SB20210824PHMHRD008488,
and SB20210993MMMHRD008470).

\subsection{Competing interest}
The authors declare no competing interests.

\subsection{Ethics approval}
Not applicable.

\subsection{Consent to participate}
Not applicable.

\subsection{Consent for publication}
Not applicable.

\subsection{Availability of data and materials}
All the data for the study is in the manuscript and in supplementary file.

\subsection{Code availability}
Not applicable.

\subsection{Authors' contributions}
S.M.S. and S.K.Y. developed the initial hypothesis. S.M.S. performed the calculations and drafted the initial manuscript.  S.M.S., S.K.Y., and B.S.M. together discussed the results. S.K.Y. and B.S.M. supervised the project. All authors contributed to the manuscript writing.

\clearpage
\printbibliography

@techreport{Baker2021,
 author =        {Baker, Matt},
 month =         {10},
 pages =         {116},
 title =         {{Defining Pathways for Realizing the Revolutionary
   Potential of High Entropy Alloys: A TMS Accelerator
   Study}},
 year =          {2021},
 doi =           {10.7449/HEApathways},
 isbn =          {9780578965949},
 url =           {https://www.tms.org/HEApathways https://www.tms.org/
  HEApathways%0Ahttps://drive.google.com/file/d/
  14TJOGnwaW4Q7MGWyXoBskWwlfA2zHpQU/view?usp=sharing},
}

@article{Yeh2006,
 author =        {Yeh, Jien-Wei},
 journal =       {Annales de Chimie Science des Mat{\'{e}}riaux},
 month =         {12},
 number =        {6},
 pages =         {633--648},
 title =         {{Recent progress in high-entropy alloys}},
 volume =        {31},
 year =          {2006},
 doi =           {10.3166/acsm.31.633-648},
 issn =          {09473580},
 url =           {http://acsm.revuesonline.com/article.jsp?articleId=9099},
}

@incollection{MURTY2019247,
 author =        {Murty, B S and Yeh, J W and Ranganathan, S and
  Bhattacharjee, P P},
 booktitle =     {High-Entropy Alloys (Second Edition)},
 edition =       {Second Edi},
 editor =        {Murty, B S and Yeh, J W and Ranganathan, S and
  Bhattacharjee, P P},
 pages =         {247--257},
 publisher =     {Elsevier},
 title =         {{13 - Applications and future directions}},
 year =          {2019},
 doi =           {https://doi.org/10.1016/B978-0-12-816067-1.00013-8},
 isbn =          {978-0-12-816067-1},
 url =           {http://www.sciencedirect.com/science/article/pii/
  B9780128160671000138},
}

@book{Reed2006,
 author =        {Reed, Roger C},
 pages =         {372},
 title =         {{The Superalloys: Fundamentals and Applications
   (Google eBook)}},
 year =          {2006},
 isbn =          {1139458639},
 url =           {http://books.google.com/books?id=SIUGcd4a-EkC&pgis=1},
}

@article{Senkov2018,
 author =        {Senkov, Oleg N. and Miracle, Daniel B. and
  Chaput, Kevin J. and Couzinie, Jean-Philippe},
 journal =       {Journal of Materials Research},
 month =         {10},
 number =        {19},
 pages =         {3092--3128},
 title =         {{Development and exploration of refractory high
   entropy alloys—A review}},
 volume =        {33},
 year =          {2018},
 doi =           {10.1557/jmr.2018.153},
 issn =          {0884-2914},
 url =           {https://www.cambridge.org/core/product/identifier/
  S088429141800153X/type/journal_article},
}

@article{Shaikh2020m,
 author =        {Shaikh, Sufyan M. and Hariharan, V.S. and
  Yadav, Satyesh K. and Murty, B.S.},
 journal =       {Intermetallics},
 month =         {12},
 number =        {May},
 pages =         {106926},
 publisher =     {Elsevier Ltd},
 title =         {{CALPHAD and rule-of-mixtures: A comparative study
   for refractory high entropy alloys}},
 volume =        {127},
 year =          {2020},
 doi =           {10.1016/j.intermet.2020.106926},
 issn =          {09669795},
 url =           {https://doi.org/10.1016/j.intermet.2020.106926 https://
  linkinghub.elsevier.com/retrieve/pii/S0966979520305112},
}

@article{Miracle2019,
 author =        {Miracle, D. B.},
 journal =       {Nature Communications},
 number =        {1},
 pages =         {1--3},
 publisher =     {Springer US},
 title =         {{High entropy alloys as a bold step forward in alloy
   development}},
 volume =        {10},
 year =          {2019},
 doi =           {10.1038/s41467-019-09700-1},
 issn =          {20411723},
 url =           {http://dx.doi.org/10.1038/s41467-019-09700-1},
}

@incollection{MURTY201913,
 author =        {Murty, B S and Yeh, J W and Ranganathan, S and
  Bhattacharjee, P P},
 booktitle =     {High-Entropy Alloys (Second Edition)},
 edition =       {Second Edi},
 editor =        {Murty, B S and Yeh, J W and Ranganathan, S and
  Bhattacharjee, P P},
 pages =         {13--30},
 publisher =     {Elsevier},
 title =         {{2 - High-entropy alloys: basic concepts}},
 year =          {2019},
 doi =           {https://doi.org/10.1016/B978-0-12-816067-1.00002-3},
 isbn =          {978-0-12-816067-1},
 url =           {http://www.sciencedirect.com/science/article/pii/
  B9780128160671000023},
}

@article{Geller2005,
 author =        {Geller, Clint B. and Smith, Richard W. and
  Hack, John E. and Saxe, Paul and Wimmer, Erich},
 journal =       {Scripta Materialia},
 month =         {2},
 number =        {3},
 pages =         {205--210},
 title =         {{A computational search for ductilizing additives to
   Mo}},
 volume =        {52},
 year =          {2005},
 doi =           {10.1016/j.scriptamat.2004.09.034},
 issn =          {13596462},
}

@article{Yang2018,
 author =        {Yang, Chaoming and Qi, Liang},
 journal =       {Physical Review B},
 month =         {1},
 number =        {1},
 pages =         {014107},
 publisher =     {American Physical Society},
 title =         {{Ab initio calculations of ideal strength and lattice
   instability in W-Ta and W-Re alloys}},
 volume =        {97},
 year =          {2018},
 doi =           {10.1103/PhysRevB.97.014107},
 issn =          {2469-9950},
 url =           {https://link.aps.org/doi/10.1103/PhysRevB.97.014107},
}

@article{Li2019,
 author =        {Li, Yu-Hao and Zhou, Hong-Bo and Liang, Linyun and
  Gao, Ning and Deng, Huiqiu and Gao, Fei and Lu, Gang and
  Lu, Guang-Hong},
 journal =       {Acta Materialia},
 month =         {12},
 pages =         {110--123},
 publisher =     {Elsevier Ltd},
 title =         {{Transition from ductilizing to hardening in
   tungsten: The dependence on rhenium distribution}},
 volume =        {181},
 year =          {2019},
 doi =           {10.1016/j.actamat.2019.09.035},
 issn =          {13596454},
 url =           {https://doi.org/10.1016/j.actamat.2019.09.035 https://
  linkinghub.elsevier.com/retrieve/pii/S1359645419306251},
}

@article{Sheikh2016,
 author =        {Sheikh, Saad and Shafeie, Samrand and Hu, Qiang and
  Ahlstr{\"{o}}m, Johan and Persson, Christer and
  Vesel{\'{y}}, Jaroslav and Z{\'{y}}ka, Jiří and
  Klement, Uta and Guo, Sheng},
 journal =       {Journal of Applied Physics},
 month =         {10},
 number =        {16},
 pages =         {164902},
 title =         {{Alloy design for intrinsically ductile refractory
   high-entropy alloys}},
 volume =        {120},
 year =          {2016},
 doi =           {10.1063/1.4966659},
 isbn =          {10.1063/1.4966659},
 issn =          {0021-8979},
 url =           {http://aip.scitation.org/doi/10.1063/1.4966659},
}

@article{Zhao2019,
 author =        {Zhao, Y. Y. and Lei, Z. F. and Lu, Z. P. and
  Huang, J. C. and Nieh, T. G.},
 journal =       {Materials Research Letters},
 number =        {8},
 pages =         {340--346},
 title =         {{A simplified model connecting lattice distortion
   with friction stress of Nb-based equiatomic
   high-entropy alloys}},
 volume =        {7},
 year =          {2019},
 doi =           {10.1080/21663831.2019.1610105},
 issn =          {2166-3831},
 url =           {https://www.tandfonline.com/doi/full/10.1080/
  21663831.2019.1610105},
}

@article{Qian2018,
 author =        {Qian, Jing and Wu, C.Y. and Fan, J.L. and Gong, H.R.},
 journal =       {Journal of Alloys and Compounds},
 month =         {3},
 pages =         {372--376},
 title =         {{Effect of alloying elements on stacking fault energy
   and ductility of tungsten}},
 volume =        {737},
 year =          {2018},
 doi =           {10.1016/j.jallcom.2017.12.042},
 issn =          {09258388},
 url =           {http://linkinghub.elsevier.com/retrieve/pii/
  S0925838817342172 https://linkinghub.elsevier.com/retrieve/
  pii/S0925838817342172},
}

@article{Zhao2019a,
 author =        {Zhao, Shijun and Osetsky, Yuri and Stocks, G. Malcolm and
  Zhang, Yanwen},
 journal =       {npj Computational Materials},
 month =         {12},
 number =        {1},
 pages =         {13},
 publisher =     {Springer US},
 title =         {{Local-environment dependence of stacking fault
   energies in concentrated solid-solution alloys}},
 volume =        {5},
 year =          {2019},
 doi =           {10.1038/s41524-019-0150-y},
 issn =          {2057-3960},
 url =           {http://dx.doi.org/10.1038/s41524-019-0150-y http://
  www.nature.com/articles/s41524-019-0150-y},
}

@article{Zaddach2013,
 author =        {Zaddach, A. J. and Niu, C. and Koch, C. C. and
  Irving, D. L.},
 journal =       {JOM},
 month =         {12},
 number =        {12},
 pages =         {1780--1789},
 title =         {{Mechanical Properties and Stacking Fault Energies of
   NiFeCrCoMn High-Entropy Alloy}},
 volume =        {65},
 year =          {2013},
 doi =           {10.1007/s11837-013-0771-4},
 isbn =          {1047-4838},
 issn =          {1047-4838},
 url =           {http://link.springer.com/10.1007/s11837-013-0771-4},
}

@article{BeyramaliKivy2017,
 author =        {Beyramali Kivy, M. and Asle Zaeem, M.},
 journal =       {Scripta Materialia},
 month =         {10},
 pages =         {83--86},
 publisher =     {Acta Materialia Inc.},
 title =         {{Generalized stacking fault energies, ductilities,
   and twinnabilities of CoCrFeNi-based face-centered
   cubic high entropy alloys}},
 volume =        {139},
 year =          {2017},
 doi =           {10.1016/j.scriptamat.2017.06.014},
 isbn =          {13596462},
 issn =          {13596462},
}

@article{Liu2018,
 author =        {Liu, S.F. and Wu, Y. and Wang, H.T. and He, J.Y. and
  Liu, J.B. and Chen, C.X. and Liu, X.J. and Wang, H. and
  Lu, Z.P.},
 journal =       {Intermetallics},
 month =         {2},
 number =        {August 2017},
 pages =         {269--273},
 title =         {{Stacking fault energy of face-centered-cubic high
   entropy alloys}},
 volume =        {93},
 year =          {2018},
 doi =           {10.1016/j.intermet.2017.10.004},
 issn =          {09669795},
}

@article{Hu2021c,
 author =        {Hu, Yong-Jie and Sundar, Aditya and Ogata, Shigenobu and
  Qi, Liang},
 journal =       {Acta Materialia},
 month =         {5},
 pages =         {116800},
 publisher =     {Elsevier Ltd},
 title =         {{Screening of generalized stacking fault energies,
   surface energies and intrinsic ductile potency of
   refractory multicomponent alloys}},
 volume =        {210},
 year =          {2021},
 doi =           {10.1016/j.actamat.2021.116800},
 issn =          {13596454},
 url =           {https://doi.org/10.1016/j.actamat.2021.116800 https://
  linkinghub.elsevier.com/retrieve/pii/S1359645421001804},
}

@article{Senkov2021d,
 author =        {Senkov, O. N. and Miracle, D. B.},
 journal =       {Scientific Reports},
 number =        {1},
 pages =         {10--13},
 publisher =     {Nature Publishing Group UK},
 title =         {{Generalization of intrinsic ductile-to-brittle
   criteria by Pugh and Pettifor for materials with a
   cubic crystal structure}},
 volume =        {11},
 year =          {2021},
 doi =           {10.1038/s41598-021-83953-z},
 isbn =          {0123456789},
 issn =          {20452322},
 url =           {https://doi.org/10.1038/s41598-021-83953-z},
}

@article{Mak2021,
 author =        {Mak, Eleanor and Yin, Binglun and Curtin, W. A.},
 journal =       {Journal of the Mechanics and Physics of Solids},
 number =        {July 2020},
 pages =         {104389},
 publisher =     {Elsevier Ltd},
 title =         {{A ductility criterion for bcc high entropy alloys}},
 volume =        {152},
 year =          {2021},
 doi =           {10.1016/j.jmps.2021.104389},
 issn =          {00225096},
 url =           {https://doi.org/10.1016/j.jmps.2021.104389},
}

@article{VandeWalle2002,
 author =        {Van de Walle, A. and Ceder, G.},
 journal =       {Reviews of Modern Physics},
 number =        {1},
 pages =         {11--45},
 title =         {{The effect of lattice vibrations on substitutional
   alloy thermodynamics}},
 volume =        {74},
 year =          {2002},
 doi =           {10.1103/RevModPhys.74.11},
 issn =          {00346861},
}

@article{Shin2007,
 author =        {Shin, Dongwon and van de Walle, Axel and Wang, Yi and
  Liu, Zi-Kui},
 journal =       {Physical Review B},
 month =         {10},
 number =        {14},
 pages =         {144204},
 title =         {{First-principles study of ternary fcc solution
   phases from special quasirandom structures}},
 volume =        {76},
 year =          {2007},
 doi =           {10.1103/PhysRevB.76.144204},
 isbn =          {1098-0121},
 issn =          {1098-0121},
 url =           {https://link.aps.org/doi/10.1103/PhysRevB.76.144204},
}

@misc{Villars2016:sm_isp_c_0902814,
 editor =        {Villars, Pierre and Okamoto, Hiroaki},
 publisher =     {Springer-Verlag Berlin Heidelberg
  {\{}{\textbackslash}{\&}{\}} Material Phases Data
  System (MPDS), Switzerland
  {\{}{\textbackslash}{\&}{\}} National Institute for
  Materials Science (NIMS), Japan},
 title =         {{Re-V Binary Phase Diagram 0-100
   at.{\{}{\textbackslash}{\%}{\}} V: Datasheet from
   ``PAULING FILE Multinaries Edition -- 2012'' in
   SpringerMaterials}},
 doi =
 {https://materials.springer.com/isp/phase-diagram/docs/c_0902814},
 url =           {https://materials.springer.com/isp/phase-diagram/docs/
  c_0902814},
}

@misc{Villars2016:sm_isp_c_0100260,
 editor =        {Villars, Pierre and Okamoto, Hiroaki},
 publisher =     {Springer-Verlag Berlin Heidelberg
  {\{}{\textbackslash}{\&}{\}} Material Phases Data
  System (MPDS), Switzerland
  {\{}{\textbackslash}{\&}{\}} National Institute for
  Materials Science (NIMS), Japan},
 title =         {{Re-Ta Binary Phase Diagram 0-100
   at.{\{}{\textbackslash}{\%}{\}} Ta: Datasheet from
   ``PAULING FILE Multinaries Edition -- 2012'' in
   SpringerMaterials}},
 doi =
 {https://materials.springer.com/isp/phase-diagram/docs/c_0100260},
 url =           {https://materials.springer.com/isp/phase-diagram/docs/
  c_0100260},
}

@misc{Villars2016:sm_isp_c_0907545,
 editor =        {Villars, Pierre and Okamoto, Hiroaki},
 publisher =     {Springer-Verlag Berlin Heidelberg
  {\{}{\textbackslash}{\&}{\}} Material Phases Data
  System (MPDS), Switzerland
  {\{}{\textbackslash}{\&}{\}} National Institute for
  Materials Science (NIMS), Japan},
 title =         {{Hf-W Binary Phase Diagram 0-100
   at.{\{}{\textbackslash}{\%}{\}} W: Datasheet from
   ``PAULING FILE Multinaries Edition -- 2012'' in
   SpringerMaterials}},
 doi =
 {https://materials.springer.com/isp/phase-diagram/docs/c_0907545},
 url =           {https://materials.springer.com/isp/phase-diagram/docs/
  c_0907545},
}

@misc{Villars2016:sm_isp_c_0904967,
 editor =        {Villars, Pierre and Okamoto, Hiroaki},
 publisher =     {Springer-Verlag Berlin Heidelberg
  {\{}{\textbackslash}{\&}{\}} Material Phases Data
  System (MPDS), Switzerland
  {\{}{\textbackslash}{\&}{\}} National Institute for
  Materials Science (NIMS), Japan},
 title =         {{W-Zr Binary Phase Diagram 0-100
   at.{\{}{\textbackslash}{\%}{\}} Zr: Datasheet from
   ``PAULING FILE Multinaries Edition -- 2012'' in
   SpringerMaterials}},
 doi =
 {https://materials.springer.com/isp/phase-diagram/docs/c_0904967},
 url =           {https://materials.springer.com/isp/phase-diagram/docs/
  c_0904967},
}

@misc{Villars2016:sm_isp_c_0100184,
 editor =        {Villars, Pierre and Okamoto, Hiroaki},
 publisher =     {Springer-Verlag Berlin Heidelberg
  {\{}{\textbackslash}{\&}{\}} Material Phases Data
  System (MPDS), Switzerland
  {\{}{\textbackslash}{\&}{\}} National Institute for
  Materials Science (NIMS), Japan},
 title =         {{Hf-V Binary Phase Diagram 0-100
   at.{\{}{\textbackslash}{\%}{\}} V: Datasheet from
   ``PAULING FILE Multinaries Edition -- 2012'' in
   SpringerMaterials}},
 doi =
 {https://materials.springer.com/isp/phase-diagram/docs/c_0100184},
 url =           {https://materials.springer.com/isp/phase-diagram/docs/
  c_0100184},
}

@misc{Villars2016:sm_isp_c_0905433,
 editor =        {Villars, Pierre and Okamoto, Hiroaki},
 publisher =     {Springer-Verlag Berlin Heidelberg
  {\{}{\textbackslash}{\&}{\}} Material Phases Data
  System (MPDS), Switzerland
  {\{}{\textbackslash}{\&}{\}} National Institute for
  Materials Science (NIMS), Japan},
 title =         {{V-Zr Binary Phase Diagram 0-100
   at.{\{}{\textbackslash}{\%}{\}} Zr: Datasheet from
   ``PAULING FILE Multinaries Edition -- 2012'' in
   SpringerMaterials}},
 doi =
 {https://materials.springer.com/isp/phase-diagram/docs/c_0905433},
 url =           {https://materials.springer.com/isp/phase-diagram/docs/
  c_0905433},
}

@misc{Villars2016:sm_isp_c_0904355,
 editor =        {Villars, Pierre and Okamoto, Hiroaki},
 publisher =     {Springer-Verlag Berlin Heidelberg
  {\{}{\textbackslash}{\&}{\}} Material Phases Data
  System (MPDS), Switzerland
  {\{}{\textbackslash}{\&}{\}} National Institute for
  Materials Science (NIMS), Japan},
 title =         {{Re-W Binary Phase Diagram 0-100
   at.{\{}{\textbackslash}{\%}{\}} W: Datasheet from
   ``PAULING FILE Multinaries Edition -- 2012'' in
   SpringerMaterials}},
 doi =
 {https://materials.springer.com/isp/phase-diagram/docs/c_0904355},
 url =           {https://materials.springer.com/isp/phase-diagram/docs/
  c_0904355},
}

@article{Zunger1990,
 author =        {Zunger, Alex and Wei, S.-H. and Ferreira, L. G. and
  Bernard, James E.},
 journal =       {Physical Review Letters},
 month =         {7},
 number =        {3},
 pages =         {353--356},
 title =         {{Special quasirandom structures}},
 volume =        {65},
 year =          {1990},
 doi =           {10.1103/PhysRevLett.65.353},
 issn =          {0031-9007},
 url =           {https://link.aps.org/doi/10.1103/PhysRevLett.65.353},
}

@article{Rice1992,
 author =        {Rice, James R.},
 journal =       {Journal of the Mechanics and Physics of Solids},
 month =         {1},
 number =        {2},
 pages =         {239--271},
 title =         {{Dislocation nucleation from a crack tip: An analysis
   based on the Peierls concept}},
 volume =        {40},
 year =          {1992},
 doi =           {10.1016/S0022-5096(05)80012-2},
 issn =          {00225096},
 url =           {https://linkinghub.elsevier.com/retrieve/pii/
  S0022509605800122},
}

@article{Waghmare1998,
 author =        {Waghmare, U. V. and Kaxiras, Efthimios and
  Bulatov, V. V. and Duesbery, M. S.},
 journal =       {Modelling and Simulation in Materials Science and
  Engineering},
 month =         {7},
 number =        {4},
 pages =         {493--506},
 title =         {{Effects of alloying on the ductility of MoSi 2
   single crystals from first-principles calculations}},
 volume =        {6},
 year =          {1998},
 doi =           {10.1088/0965-0393/6/4/013},
 issn =          {0965-0393},
 url =           {https://iopscience.iop.org/article/10.1088/0965-0393/6/4/
  013},
}

@article{Schuh2018,
 author =        {Schuh, B. and V{\"{o}}lker, B. and Todt, J. and
  Schell, N. and Perri{\`{e}}re, L. and Li, J. and
  Couzini{\'{e}}, J. P. and Hohenwarter, A.},
 journal =       {Acta Materialia},
 pages =         {201--212},
 title =         {{Thermodynamic instability of a nanocrystalline,
   single-phase TiZrNbHfTa alloy and its impact on the
   mechanical properties}},
 volume =        {142},
 year =          {2018},
 doi =           {10.1016/j.actamat.2017.09.035},
 isbn =          {1359-6454},
 issn =          {13596454},
}

@article{Juan2016,
 author =        {Juan, Chien Chang and Tsai, Ming Hung and
  Tsai, Che Wei and Hsu, Wei Lin and Lin, Chun Ming and
  Chen, Swe Kai and Lin, Su Jien and Yeh, Jien Wei},
 journal =       {Materials Letters},
 pages =         {200--203},
 publisher =     {Elsevier},
 title =         {{Simultaneously increasing the strength and ductility
   of a refractory high-entropy alloy via grain
   refining}},
 volume =        {184},
 year =          {2016},
 doi =           {10.1016/j.matlet.2016.08.060},
 issn =          {18734979},
 url =           {http://dx.doi.org/10.1016/j.matlet.2016.08.060},
}

@article{ChenPeierlsAlloy,
 author =        {Chen, S. Y. and Wang, L. and Li, W. D. and Tong, Y. and
  Tseng, K. K. and Tsai, C. W. and Yeh, J. W. and
  Ren, Y. and Guo, W. and Poplawsky, J. D. and
  Liaw, P. K.},
 journal =       {Materials Research Letters},
 title =         {{Peierls barrier characteristic and anomalous strain
   hardening provoked by dynamic-strain-aging
   strengthening in a body-centered-cubic high-entropy
   alloy}},
 doi =           {10.1080/21663831.2019.1658233},
 url =           {https://www.tandfonline.com/doi/full/10.1080/
  21663831.2019.1658233?af=R&utm_source=researcher_app&
  utm_medium=referral&
  utm_campaign=RESR_MRKT_Researcher_inbound},
}

@article{Casillas-Trujillo2020,
 author =        {Casillas-Trujillo, Luis and Jansson, Ulf and
  Sahlberg, Martin and Ek, Gustav and
  Nyg{\aa}rd, Magnus M and S{\o}rby, Magnus H and
  Hauback, Bjørn C and Abrikosov, Igor A and
  Alling, Björn},
 journal =       {Physical Review Materials},
 month =         {12},
 number =        {12},
 pages =         {123601},
 publisher =     {American Physical Society},
 title =         {{Interstitial carbon in bcc HfNbTiVZr high-entropy
   alloy from first principles}},
 volume =        {4},
 year =          {2020},
 doi =           {10.1103/PhysRevMaterials.4.123601},
 issn =          {2475-9953},
 url =           {https://doi.org/10.1103/PhysRevMaterials.4.123601 https://
  link.aps.org/doi/10.1103/PhysRevMaterials.4.123601},
}

@article{Soni2018,
 author =        {Soni, V. and Senkov, O. N. and Gwalani, B. and
  Miracle, D. B. and Banerjee, R.},
 journal =       {Scientific Reports},
 number =        {1},
 pages =         {1--10},
 publisher =     {Springer US},
 title =         {{Microstructural Design for Improving Ductility of An
   Initially Brittle Refractory High Entropy Alloy}},
 volume =        {8},
 year =          {2018},
 doi =           {10.1038/s41598-018-27144-3},
 issn =          {20452322},
 url =           {http://dx.doi.org/10.1038/s41598-018-27144-3},
}

@article{Senkov2011,
 author =        {Senkov, O.N. and Scott, J.M. and Senkova, S.V. and
  Miracle, D.B. and Woodward, C.F.},
 journal =       {Journal of Alloys and Compounds},
 month =         {5},
 number =        {20},
 pages =         {6043--6048},
 title =         {{Microstructure and room temperature properties of a
   high-entropy TaNbHfZrTi alloy}},
 volume =        {509},
 year =          {2011},
 doi =           {10.1016/j.jallcom.2011.02.171},
 isbn =          {2041-1723 (Electronic) 2041-1723 (Linking)},
 issn =          {09258388},
 url =           {https://linkinghub.elsevier.com/retrieve/pii/
  S0925838811005536},
}

@article{Senkov2015b,
 author =        {Senkov, O. N. and Semiatin, S. L.},
 journal =       {Journal of Alloys and Compounds},
 month =         {11},
 pages =         {1110--1123},
 publisher =     {Elsevier},
 title =         {{Microstructure and properties of a refractory
   high-entropy alloy after cold working}},
 volume =        {649},
 year =          {2015},
 doi =           {10.1016/j.jallcom.2015.07.209},
 issn =          {09258388},
 url =           {https://www.sciencedirect.com/science/article/pii/
  S0925838815306071},
}

@article{Ren2018,
 author =        {Ren, Chai and Fang, Z. Zak and Koopman, Mark and
  Butler, Brady and Paramore, James and
  Middlemas, Scott},
 journal =       {International Journal of Refractory Metals and Hard
  Materials},
 number =        {January},
 pages =         {170--183},
 publisher =     {Elsevier},
 title =         {{Methods for improving ductility of tungsten - A
   review}},
 volume =        {75},
 year =          {2018},
 doi =           {10.1016/j.ijrmhm.2018.04.012},
 issn =          {22133917},
 url =           {https://doi.org/10.1016/j.ijrmhm.2018.04.012},
}

@article{Walle2003,
 author =        {van de Walle, A. and Asta, M and Ceder, G.},
 journal =       {Calphad},
 month =         {12},
 number =        {4},
 pages =         {539--553},
 title =         {{The Alloy Theoretic Automated Toolkit: A User
   Guide}},
 volume =        {26},
 year =          {2002},
 doi =           {10.1016/S0364-5916(02)80006-2},
 isbn =          {0364-5916},
 issn =          {03645916},
}

@article{VanDeWalle2013,
 author =        {Van De Walle, A. and Tiwary, P. and De Jong, M. and
  Olmsted, D. L. and Asta, M. and Dick, A. and Shin, D. and
  Wang, Y. and Chen, L. Q. and Liu, Z. K.},
 journal =       {Calphad: Computer Coupling of Phase Diagrams and
  Thermochemistry},
 pages =         {13--18},
 publisher =     {Elsevier},
 title =         {{Efficient stochastic generation of special
   quasirandom structures}},
 volume =        {42},
 year =          {2013},
 doi =           {10.1016/j.calphad.2013.06.006},
 issn =          {03645916},
}

@article{Xu2011,
 author =        {Xu, Yibin and Yamazaki, Masayoshi and
  Villars, Pierre},
 journal =       {Japanese Journal of Applied Physics},
 number =        {11 PART 2},
 title =         {{Inorganic materials database for exploring the
   nature of material}},
 volume =        {50},
 year =          {2011},
 doi =           {10.1143/JJAP.50.11RH02},
 issn =          {00214922},
}

@article{Momma:db5098,
 author =        {Momma, Koichi and Izumi, Fujio},
 journal =       {Journal of Applied Crystallography},
 month =         {12},
 number =        {6},
 pages =         {1272--1276},
 title =         {{VESTA 3 for three-dimensional visualization of
   crystal, volumetric and morphology data}},
 volume =        {44},
 year =          {2011},
 doi =           {10.1107/S0021889811038970},
 issn =          {0021-8898},
}

@article{Kresse1993,
 author =        {Kresse, G. and Hafner, J.},
 journal =       {Physical Review B},
 number =        {1},
 pages =         {558--561},
 title =         {{Ab initio molecular dynamics for liquid metals}},
 volume =        {47},
 year =          {1993},
 doi =           {10.1103/PhysRevB.47.558},
 issn =          {01631829},
}

@article{Blochl1994,
 author =        {Bl{\"{o}}chl, P. E.},
 journal =       {Physical Review B},
 number =        {24},
 pages =         {17953--17979},
 title =         {{Projector augmented-wave method}},
 volume =        {50},
 year =          {1994},
 doi =           {10.1103/PhysRevB.50.17953},
 issn =          {01631829},
}

@article{Kresse1996,
 author =        {Kresse, G and Furthm{\"{u}}ller, J},
 journal =       {Physical Review B},
 month =         {10},
 number =        {16},
 pages =         {11169--11186},
 publisher =     {American Physical Society},
 title =         {{Efficient iterative schemes for ab initio
   total-energy calculations using a plane-wave basis
   set}},
 volume =        {54},
 year =          {1996},
 doi =           {10.1103/PhysRevB.54.11169},
}

@article{Perdew1996,
 author =        {Perdew, John P. and Burke, Kieron and
  Ernzerhof, Matthias},
 journal =       {Physical Review Letters},
 number =        {18},
 pages =         {3865--3868},
 title =         {{Generalized gradient approximation made simple}},
 volume =        {77},
 year =          {1996},
 doi =           {10.1103/PhysRevLett.77.3865},
 issn =          {10797114},
}

@article{Joubert1999,
 author =        {Joubert, D.},
 journal =       {Physical Review B - Condensed Matter and Materials
  Physics},
 number =        {3},
 pages =         {1758--1775},
 title =         {{From ultrasoft pseudopotentials to the projector
   augmented-wave method}},
 volume =        {59},
 year =          {1999},
 doi =           {10.1103/PhysRevB.59.1758},
 issn =          {1550235X},
}

@article{Methfessel1989,
 author =        {Methfessel, M. and Paxton, A. T.},
 journal =       {Physical Review B},
 number =        {6},
 pages =         {3616--3621},
 title =         {{High-precision sampling for Brillouin-zone
   integration in metals}},
 volume =        {40},
 year =          {1989},
 doi =           {10.1103/PhysRevB.40.3616},
 issn =          {01631829},
}

@article{Hu2019,
 author =        {Monkhorst, Hendrik J. and Pack, James D.},
 journal =       {Physical Review B},
 month =         {6},
 number =        {12},
 pages =         {5188--5192},
 title =         {{Special points for Brillouin-zone integrations}},
 volume =        {13},
 year =          {1976},
 doi =           {10.1103/PhysRevB.13.5188},
 issn =          {0556-2805},
 url =           {https://link.aps.org/doi/10.1103/PhysRevB.13.5188},
}

\end{document}